\documentclass[%
 reprint,
amsmath,amssymb,
aps,
]{revtex4-2}

\usepackage{graphicx}
\usepackage{dcolumn}
\usepackage{bm}
\usepackage{multirow}

\usepackage{xcolor}

\begin{document}


\title{Opinion formation under mass media influence on the Barabasi-Albert network}

\author{Ramacos Fardela}
\email{ramacosfardela@sci.unand.ac.id}
\thanks{corresponding author}%
\author{Zulfi Abdullah}%
 \email{zulfi@sci.unand.ac.id}
\affiliation{%
Department of Physics, Universitas Andalas, Padang, Limau Manis, 25163, Indonesia
}%

\author{Roni Muslim}
\email{roni.muslim@brin.go.id}
\affiliation{%
 Research Center for Quantum Physics,  National Research and Innovation Agency (BRIN),
South Tangerang, 15314, Indonesia
}%

\date{\today}

\begin{abstract}
We study numerically the dynamics of opinion formation under the influence of mass media using the $q$-voter model on a Barabasi-Albert network. We investigate the scenario where a voter adopts the mass media's opinion with a probability $p$ when there is no unanimity among a group of $q$ agents. Through numerical simulation, we identify a critical probability threshold, $p_t$, at which the system consistently reaches complete consensus. This threshold probability $p_t$ decreases as the group size $q$ increases, following a power-law relation $p_t \propto q^{\gamma}$ with $\gamma \approx -1.187$. Additionally, we analyze the system's relaxation time, the time required to reach a complete consensus state. This relaxation time increases with the population size $N$, following a power-law $\tau \propto N^{\nu}$, where $\nu \approx 1.093$. Conversely, an increase in the probability $p$ results in a decrease in relaxation time following a power-law relationship $\tau \propto p^{\delta}$, with $\delta \approx -0.596$. The value of the exponent \( \nu \) is similar to the exponents obtained in the voter and $q$-voter models across various network topologies.
\end{abstract}

\maketitle


\section{Introduction}

The principles of statistical physics have been successfully applied to understanding various social phenomena, thus giving rise to the interdisciplinary field known as sociophysics \cite{galam1999application, galam2012socio, castellano2009statistical, sen2014sociophysics, schweitzer2018sociophysics}. One of the main areas of study within this field, opinion dynamics, explores interactions among individuals, which can often be illustrated through mathematical models. This area of research investigates how individual opinions are formed and evolve through interactions, frequently leading to the emergence of collective behaviors that mirror social phenomena \cite{castellano2009statistical, noorazar2020classical, noorazar2020recent}.

One of the well-known opinion dynamics models in sociophysics is the $q$-voter model, an extension of the voter model. The $q$-voter model explains how a group of agents with the same opinion (unanimous), called agents of size $q$ ($q$-sized agent), can influence other agents, called voters. If $q$-sized agent is not unanimous, the voter can still change his opinion with a certain probability \cite{castellano2009nonlinear}. From a statistical physics perspective, the $q$-voter model can show various interesting statistical physics properties such as the emergence of continuous and discontinuous phase transition phenomena, as well as universality behavior with certain critical exponential values \cite{nyczka2012phase, chmiel2015phase, abramiuk2019independence, nowak2021discontinuous}.

Within the context of social media or mass media, the study of opinion dynamics models is particularly compelling. This interest is mainly due to the profound influence that mass media—including television, radio, and online platforms such as Facebook—exerts on individual opinions. Individuals may adopt or change their views based on the information these mass media, for instance, when choosing a political candidate during elections. A multitude of studies has explored the effects of external forces, like mass media, on the opinion dynamics \cite{mazzitello2007effects,candia2008mass,rodriguez2010effects,pabjan2008model,sousa2008effects, crokidakis2012effects,gonzalez2007information,gonzalez2005nonequilibrium,martins2010mass,quattrociocchi2014opinion,pineda2015mass, colaiori2015interplay,sirbu2017opinion,li2020effect, freitas2020imperfect, tiwari2021modeling, helfmann2023modelling}. These investigations have yielded interesting findings, investigating how mass media influences the formation and evolution of public opinion, facilitates the shift from a state of pluralism to consensus under its influence, and accelerates the spread of opinions through its intervention \cite{pineda2015mass,colaiori2015interplay,li2020effect,gimenez2021opinion, ishii2021sociophysics}.

The phenomena of phase changes within opinion dynamics models provide insights into societal transformations, such as shifts towards polarization to consensus, mirroring transitions seen in physical systems. Interestingly, the influence of mass media can modify or entirely negate the existence of two ordered states \cite{crokidakis2012effects, azhari2023external, muslim2024mass}. Specifically, when reaching a certain threshold probability, denoted as $p_t$, a complete consensus (all agents having the same opinion ''up") among agents emerges, signifying the absence of two consensus states. This phenomenon emphasizes the role of external influences, such as mass media, in forming collective opinion dynamics. However, in the context of real social interactions, selecting interaction networks is crucial for a deeper understanding of the dynamics involved. Therefore, examining models based on the Barabási-Albert network is more insightful and can capture real social interactions more accurately than a complete graph \cite{newman2018networks}

This study delves into the effects of mass media on the dynamics of opinion formation based on the $q$-voter model on a Barabasi-Albert network, following the scenario on our prior works \cite{azhari2023external, muslim2024mass}. Through Monte Carlo simulations, we examine the influence of mass media on the evolution of opinions. Our results demonstrate that mass media presence can drive the system towards a uniform state (complete consensus) at a certain probability threshold $p_t$. Notably, even with a low initial opinion ``up" $c_0 = 0.01$, at $p_t$, the system uniformly aligns to the ``up" opinion ($c = 1$). The probability threshold $p_t$ significantly decreases as the $q$-sized agent increases, obeying a power-law $p_t(q) \propto q^{\alpha}$ with $\alpha \approx -1.187$. Additionally, we obtain a power-law relationship between the relaxation time $\tau$ and the population size $N$, namely $\tau(N) \propto N^{\nu}$, where $\nu \approx 1.093$. The relaxation time $\tau$ also decreases as $p$ increases and follows a power-law relation $\tau(p) \propto p^{\delta}$, where $\delta \approx -0.560$.

\section{Model and Method}
We examine the influence of mass media on public opinion using the $q$-voter model within a Barabasi-Albert network. The Barabasi-Albert network is a scale-free network characterized by preferential attachment, where new nodes are more likely to connect to well-established nodes \cite{albert2002statistical}. This model mirrors the ``rich get richer" phenomenon observed in real social networks, where popular individuals or content attract more connections, resulting in a highly skewed distribution of connections. In this network, agents are randomly assigned to the network's nodes and interact with their nearest neighbors, who are directly connected through the network's edges or links, which represent social connections.

\begin{figure}[tb!]
    \centering
    \includegraphics[width = \linewidth]{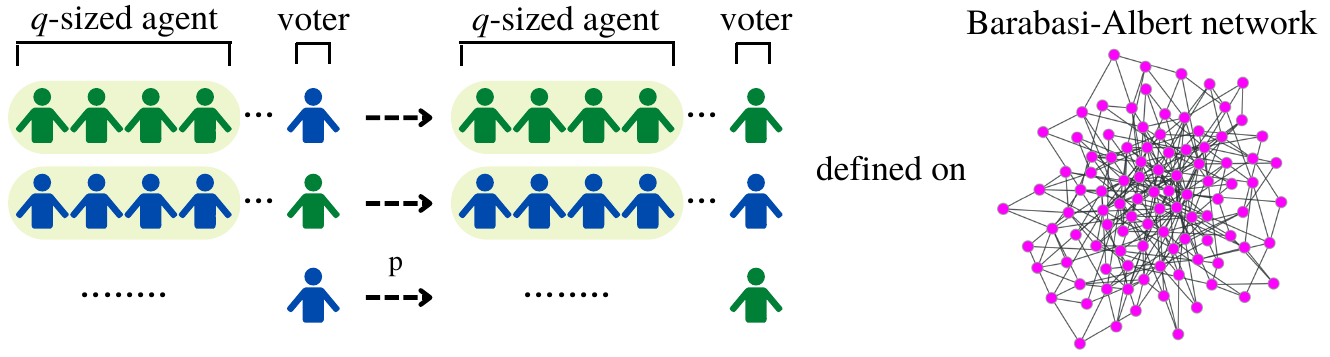}
    \caption{An illustration of the $q$-voter model with mass media influence on the Barabasi-Albert network is presented. The two upper parts of the figure present two scenarios following the original $q$-voter model, where a voter adopts the opinion of a group of size \(q\) if the group shares the same opinion. The lower part illustrates a mass media scenario where the voter changes their opinion with a probability of \(p\) when the group of size \(q\) does not have a unanimous opinion. The green and blue colors represent the two differing opinions or states.}
    \label{Fig01}
\end{figure}

Within the original $q$-voter model framework, a group of agents, designated as $S_q$, is randomly selected from the population to potentially influence another agent, identified as the voter, designated as $S_{q+1}$. If the selected group $S_q$ exhibits unanimity in opinion, the voter $S_{q+1}$ adopts this unanimous opinion. Conversely, if the group is not unanimous, the voter $S_{q+1}$ may still change their opinion with a certain probability $\epsilon$ \cite{castellano2009nonlinear}. This study follows the original model outlined in Ref.~\cite{castellano2009nonlinear}, setting $\epsilon = 0$, similar to several works in Refs.~\cite{nyczka2013anticonformity, civitarese2021external, azhari2023external, muslim2024mass}. Furthermore, we introduce a scenario involving mass media influence. If the group's opinion is not homogeneous, the voter adopts the mass media's opinion with a probability $p$, as illustrated in the bottom scenario of Fig.~\ref{Fig01}.

In this model, each agent is assumed to have two possible opinions, \( \pm 1 \). The mass media influence prompts the voter to adopt the opinion \( +1 \), specifically changing \( S_{q+1}(t) = -1 \) to \( S_{q+1}(t+1) = +1 \), when the opinion of agents \( S_{q} \) do not share the same opinion. Consequently, mass media influence drives the population towards a complete consensus, where all individuals hold the opinion $+1$. In real-world social scenarios, this model can reflect situations involving small groups of individuals, where internal influences from $q$-sized groups (such as familial, social, and professional circles) outweigh the mass media influence. Thus, we set the probability of voters changing their opinions to $1$ when the $q$-sized agents share the same opinion, as depicted in the two top scenarios in Fig.~\ref{Fig01}. Additionally, we restrict the probability \( p \) to assume values from 0 to 1 when under the influence of mass media, with 0 indicating the absence of mass media influence and 1 representing the strongest possible influence.

\textcolor{black}{More specifically, the algorithm of the model can be summarized as follows:}
\textcolor{black}{\begin{enumerate}
    \item The initial state of the system is prepared with an initial proportion of positive opinions \( c_0 = [0,1] \), where \( c_0 = 0 \) indicates that no agents have the \( +1 \) opinion initially, and \( c_0 = 1 \) indicates that all agents have the \( +1 \) opinion initially.
    \item At time \( t \), a voter (\( S_{q+1} \)) is randomly selected, and a group of \( q \) agents (\( S_{q} \)), who are the nearest neighbors (directly connected), are also randomly selected.
    \item If the group of \( q \) agents shares the same opinion, the voter will adopt the opinion of this group. If the group of \( q \) agents does not share the same opinion, then with probability \( p = [0,1]\), the voter adopts the opinion of the mass media, which is \( +1 \).
\end{enumerate}}

To analyze the average public opinion within this model, we define the order parameter $m$ as follows:
\begin{equation}\label{eq:magnetization}
    m = \sum_{i=1}^{N} \dfrac{S_i}{N}.
\end{equation}
This order parameter is analogous to magnetization in the Ising model and \textcolor{black}{also can be written as $m = 2\,c-1$}. Here, $N$ denotes the total number of agents (or the total number of network nodes), and $S_i$ represents the $i$th agent's opinion. The order parameter's value $m = \pm 1$ indicates a complete consensus situation where all agents have the same opinion.

In each step of the Monte Carlo simulation (time), the population opinion can undergo one of three possible changes: increase by \( 1/N \), decrease by \( -1/N \), or remain unchanged. \textcolor{black}{In each Monte Carlo step, a voter and a group of \( q \) agents, who are the nearest neighbors of the voter, are randomly selected and then interact based on the algorithm described above. The system's opinion is then updated and recorded. This process continues until a consensus state is reached}. Our investigation focuses on determining the existence of a separator point in the order parameter $m$ plotted against the probability $p$ of mass media influence, distinguishing between two distinct complete consensus states with $m = +1$ and $m = -1$. This inquiry is particularly relevant when an initial fraction of opinion favoring $+1$, denoted by $c_0$, is very small, for example, $c_0 = 0.01$.

Suppose such a separator point cannot be identified at a given probability $p$. In that case, we designate this probability as the threshold probability $p_t$, signifying the absence of a scenario where public opinion converges to two distinct consensus states. In other words, at the threshold probability $p_t$, public opinion will never reach a complete consensus where all agents hold the opinion $-1$. This methodology aligns with previous research on the impact of mass media in the Sznajd model on a 2-D square lattice \cite{crokidakis2012effects}, as well as in the majority-rule and $q$-voter models on a complete graph \cite{azhari2023external}, and specifically, the $q$-voter model on the complete graph \cite{muslim2024mass}. The key takeaway is that mass media influence effectively eliminates the critical separator point when $p \geq p_t$. For instance, even with a minimal initial opinion fraction of $c_0 = 0.01$, the system transitions to a uniform state with $m = +1$ for $p \geq p_t$. Without mass media influence ($p = 0$), a separator point at $c = c_s = 0.5$ traditionally marks the transition to uniform states of $m = -1$ for $c_0 < 0.5$ and $m = +1$ for $c_0 > 0.5$ in large-scale systems \cite{stauffer2000generalization}.

To investigate the scaling behavior within this model, we utilized standard finite-size scaling analysis to delineate the scaling parameters associated with the order parameter $m$. The formula for the standard finite-size scaling relationship is expressed as $m(c,N) = N^{-\gamma_1} \phi((c-c_s)N^{-\gamma_2})$
where $c = c_s + \gamma_1 N^{-\gamma_2} $ \cite{sousa2008effects}. Here, $\gamma_1$ and $\gamma_2$ are the scaling parameters that facilitate the optimal data collapse, and $\phi$ is the dimensionless function governing the scaling behavior. The critical separator point, $c_s$, is determined through the intersection of the order parameter $m$ with the initial opinion $c_0$. 

One of the interesting parameters within the Monte Carlo simulation is the relaxation time, or consensus time, denoted as $\tau$. The relaxation time is required for all agents to reach a homogeneous state, where all agents adopt the same opinion. In this study, we investigate the effect of mass media influence on the relaxation time $\tau$ and examine how the relaxation time depends on the population size $N$ of the system. This analysis can provide insights into how quickly a population can achieve consensus under varying degrees of mass media influence and different population sizes.

\section{Result and Discussion}
\begin{figure}[tb]
    \centering
    \includegraphics[width = \linewidth]{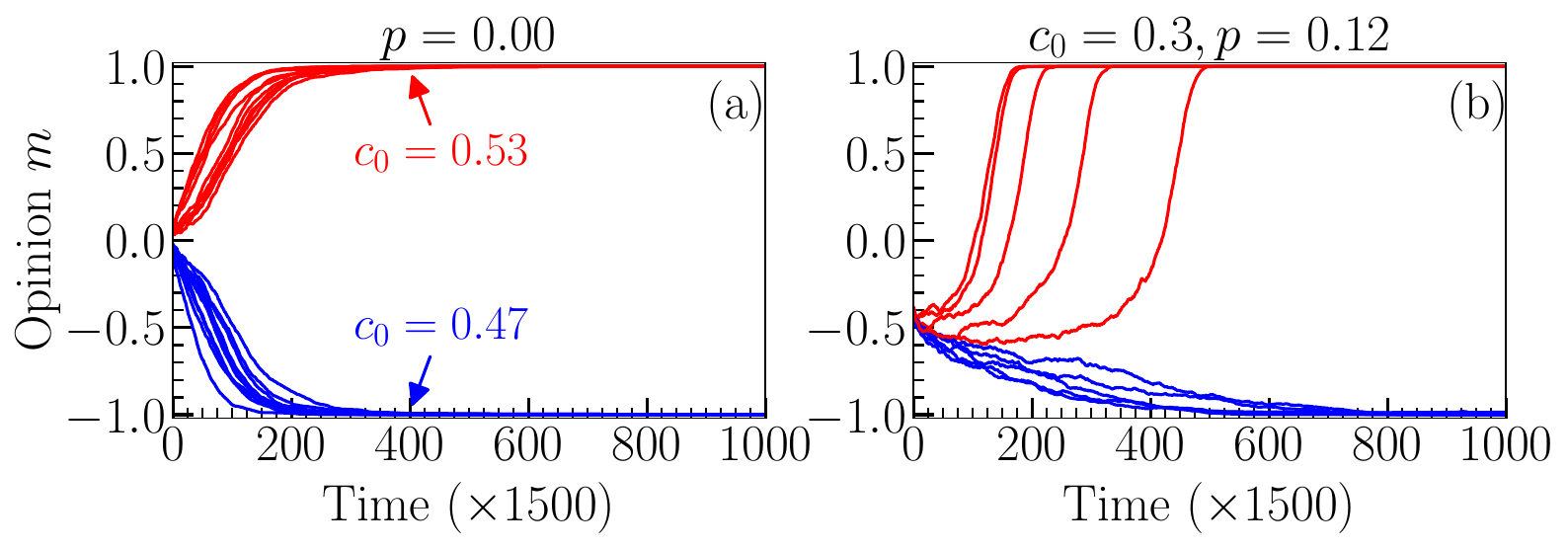}
    \caption{The comparison of the evolution of the average public opinion \(m\) based on the $q$-voter model on the Barabási-Albert network illustrates contrasting effects of mass media influence. Without mass media [panel (a), \( p = 0 \)], \( m \) reaches a complete consensus with \( m = 1 \) for \( c_0 > 0.5 \) (red color) and \( m = -1 \) for \( c_0 < 0.5 \) (blue color), indicating that the separator point in this case is at \( c = 0.5 \). With mass media influence [panel (b), \( p = 0.12 \)], several samples evolve to \( m = 1 \) (red color) even for \( c_0 = 0.3 < 0.5 \), indicating that the separator point shifts from \( c = 0.5 \) and depending on \( p \). The network size is \( N = 5000 \) and \( q = 2 \).}
    \label{fig:mcs}
\end{figure}

To investigate the impact of mass media on public opinion, we begin by analyzing the evolution of the order parameter $m$, as defined in Eq.~\eqref{eq:magnetization}, in scenarios both including and excluding the influence probability $p$. We start with the case where $q = 2$ and the initial opinion fraction is less than $0.5$, examining conditions without mass media influence ($p = 0$) and with mass media influence at a probability of $p = 0.12$. Figure~\ref{fig:mcs} (a) illustrates that in several instances, the system achieved a unanimous opinion (complete consensus), with $m = 1$ for an initial opinion fraction of $c_0 = 0.53$ and $m = -1$ for $c_0 = 0.47$. This outcome suggests that the separator point, which separates the two unanimous opinion states at equal probabilities, is $c = c_s = 0.5$. As depicted in panel (b), the presence of mass media influence shifts the separator point $c_s$ away from $c = 0.5$, observed in the absence of mass media ($p = 0$), to a specific value. Thus, even with an initial opinion fraction $c_0$ below 0.5, numerous samples could achieve a consensus at $m = 1$. Therefore, this model's separator point $c_s$ depends on the probability $p$.

\begin{figure}[t]
    \centering
    \includegraphics[width = \linewidth]{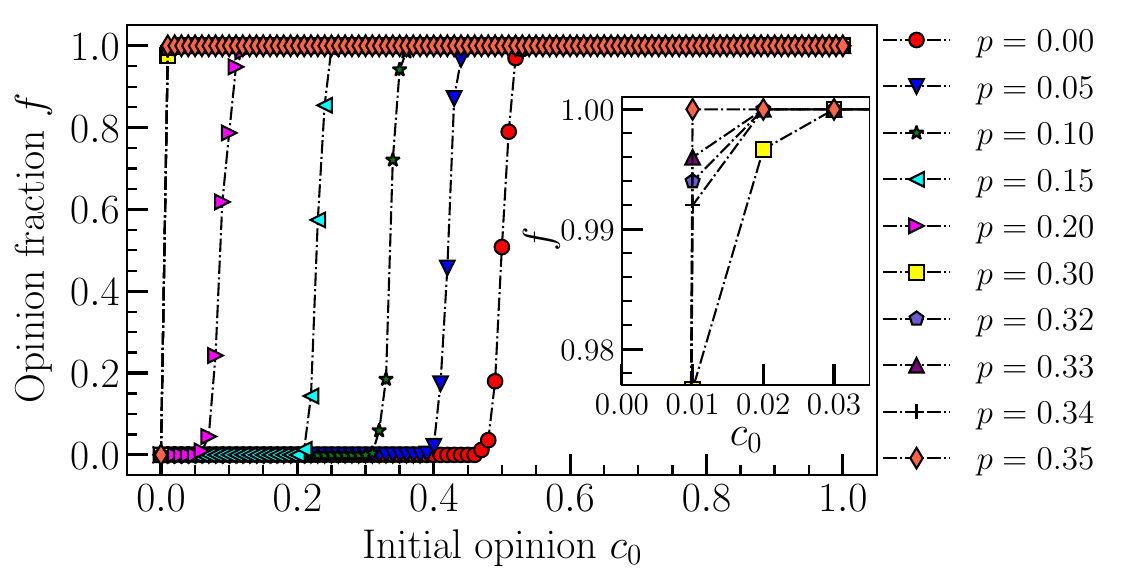}
    \caption{The numerical results illustrate how the average opinion fraction \(f\) varies to the initial opinion \(c_0\) across different levels of mass media influence probability \(p\), for \(q = 2\). Clearly, when \(p = 0.35\), the fraction \(f\) reaches 1 for a minimal initial opinion of \(c_0 = 0.01\). This result indicates that the critical separation point vanishes when \(p\) equals the threshold value \(p_t = 0.35\). Each data point is derived from an average of 500 samples and is based on a population size of \(N = 5000\).}
    \label{fig:fraction_m}
\end{figure}

To more clearly understand the role of mass media within our model, we varied the initial opinion fraction $c_0$ across different probabilities of mass media influence $p$ to calculate the average order parameter $m$. Our goal was to ascertain the presence of a separator point in the model, emphasizing a low initial opinion fraction, such as $c_0 = 0.01$, where we expected to observe an order parameter $m$ consistently below $1$ for all values of $p$. We analyzed a large sample size of $N = 5000$ to obtain reliable results, averaging outcomes over $500$ samples for each data point. In a scenario where $q = 2$, the numerical results for the order parameter $m$ are exhibited in Fig.~\ref{fig:fraction_m}. Notably, at $p = 0$ (without mass media influence), the separation point occurs at $c = 0.5$, indicating that the system has an equal probability of achieving a complete consensus of either $+1$ or $-1$. However, with $p \neq 0$, the separation point no longer sits at $c = 0.5$ but shifts to a different location, depending on the probability $p$. 

Figure~\ref{fig:fraction_m} (see the inset graph for more details) shows that at $c_0 = 0.01$, the system reaches a complete consensus ($m = 1$) for $p = 0.35$. In other words, we consistently observed that the average public opinion moves towards a complete consensus, with all agents sharing the same positive opinion, when the mass media influence $p \geq 0.35$. Therefore, at $p \geq 0.35$, the separation point $c_s$, which delineates the probability of public opinion splitting into two distinct consensuses ($m = 1$ and $m = -1$), is no longer evident. Thus, we can say that the threshold probability of mass media influence required to eliminate the separator point $c_s$ is $p_t = 0.35$.

\begin{figure}[tb]
    \centering
    \includegraphics[width = \linewidth]{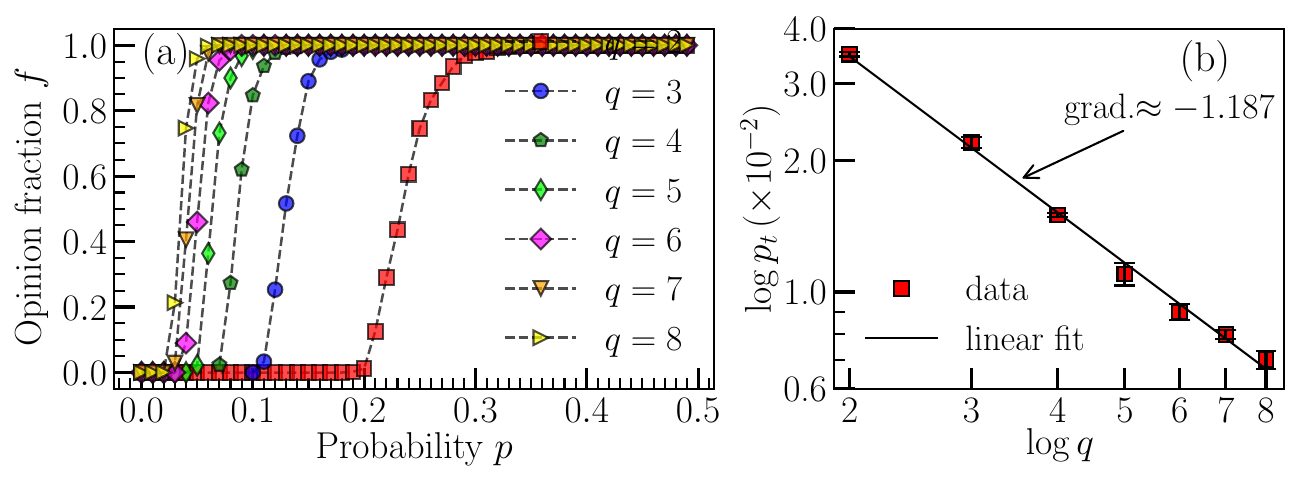}
    \caption{(a) Numerical results for the opinion fraction \(f\) versus probability \(p\) at \(c_0 = 0.01\) for various \(q\)-size agents. A value \(p = p_t\) exists that sets the value of \(f = 1\), indicating that all agents share the same opinion at this \(p\) value. (b) A log-log plot of the threshold probability \(p_t\) versus \(q\)-size agent shows a power-law relationship (see the inset for the normal plot). Each data point is based on the average of 500 samples and involves a population size of \(N = 5000\).}
    \label{fig:qvspt}
\end{figure}

To determine the threshold value $p_t$ in the model incorporating agents of various $q$ sizes, we examined the relationship between the opinion fraction $f$ and the probability $p$ for an initial opinion fraction of $c_0 = 0.01$. The results are exhibited in Fig.~\ref{fig:qvspt} (a). We observed that the threshold value $p_t$ varies according to the size of the $q$-sized agent, with $p_t$ decreasing as $q$ increases [see Fig.~\ref{fig:qvspt} (b)]. This trend is consistent with our previous results for the \( q \)-voter model defined on a complete graph \cite{muslim2024mass}.  Intriguingly, as depicted by the log-log plot in panel (b), the relationship between the threshold value $p_t$ and the size of $q$ adheres to a power-law, which can be expressed as follows:
\begin{equation}
    p_t(q) \propto q^{\alpha},
\end{equation}
where the value of $\alpha \approx -1.187$. This power-law relationship indicates that the threshold probability $p_t$ decreases as the group size $q$ increases, suggesting that larger groups of agents require a lower probability of mass media influence to achieve consensus. This $\alpha$ value slightly differs from that obtained in previous studies, where $\alpha \approx -1$ for the model analyzed on a complete graph \cite{muslim2024mass}.

\begin{figure}[tb]
    \centering
    \includegraphics[width = 0.75\linewidth]{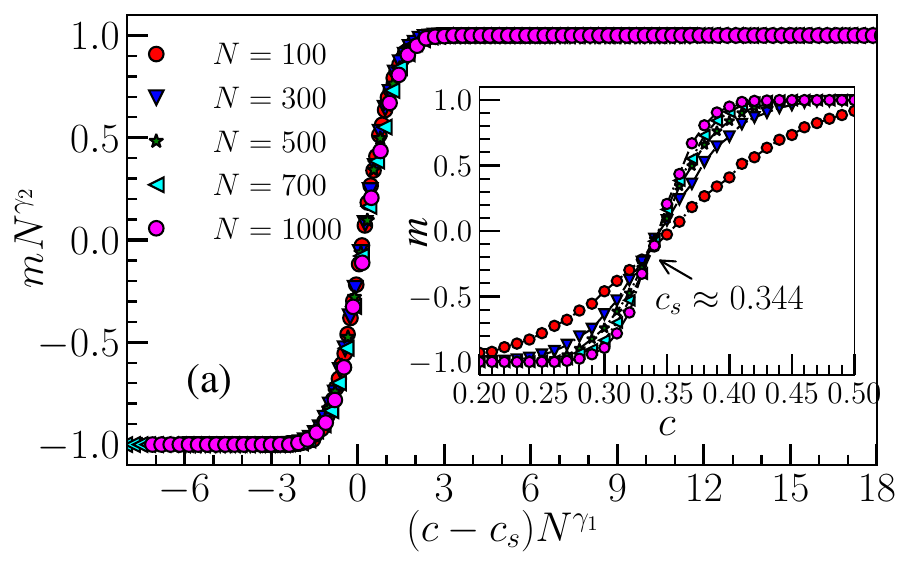}
    \caption{A scaling plot of the order parameter \(m \) (public opinion) for \(q = 2\) and a probability of \(p = 0.1\) shows how the data aligns under the scaling parameters \(\gamma_1 \approx 0.02\) and \(\gamma_2 \approx 0.498\). The separation point \(c_s \approx 0.334\) is identified through the intersection of the lines plotting \(m\) against \(c\), as depicted in the inset graph.}
    \label{fig:subfig_a}
\end{figure}

As previously mentioned, we utilize standard finite-size scaling analysis to obtain the separation point $c_s$ and determine the scaling parameters $\gamma_1$ and $\gamma_2$, facilitating data alignment across different population sizes $N$. For a chosen probability $p = 0.1$, less than the threshold $p_t(q=2)$, we varied the population size $N$, with each data point representing an average of over 5000 samples. The intersection point of the curves representing the order parameter $m$, identified at $c = c_s \approx 0.344$ [see the inset of Fig.~\ref{fig:subfig_a}], marks the model's separation point. This point divides the system into two complete consensus states: $m = -1$ for $c < c_s$ and $m = 1$ for $c > c_s$.

\begin{figure}[tb]
    \centering
    \includegraphics[width = 0.75\linewidth]{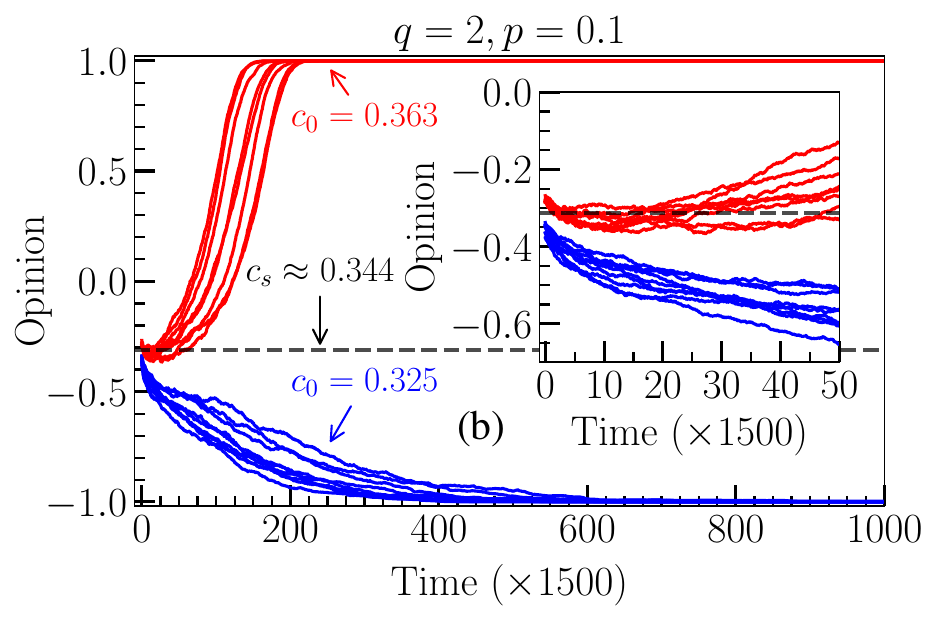}
    \caption{The temporal dynamics of the order parameter \(m\), with the initial opinion \(c_0\) taken close to the separator point \(c_s\), reveal distinct outcomes for \(c_0 = 0.352 < c_s\) and \(c_0 = 0.363 > c_s\). All samples evolve towards a homogeneous state, indicated by \(m = +1\) (in red) for \(c_0 > c_s\) and \(m = -1\) (in blue) for \(c_0 < c_s\). The population size is \(N = 5000\).}
    \label{fig:subfig_b}
\end{figure}

By fine-tuning the scaling parameters, we obtained the best collapse of data, resulting in $\gamma_1 \approx 0.002$ and $\gamma_2 \approx 0.498$, values that are similar to those found in our previous studies \cite{azhari2023external, muslim2024mass}. These scaling parameters are considered universal, meaning they work across all population sizes $N$ and probabilities $p \leq p_t(q)$ for all values of $q$-sized agents. Additionally, we explored the behavior of the order parameter $m$ for $q = 2$, starting with an opinion fraction $c_0$ near the separation point $c_s \approx 0.334$ for a large population size of $N = 5000$, as shown in Fig.~\ref{fig:subfig_b}. The analysis reveals that for $c < c_s$, all samples move towards a homogeneous state with the order parameter $m = -1$, while for $c > c_s$, all samples move towards $m = 1$.

\begin{figure}[tb]
    \centering
    \includegraphics[width = \linewidth]{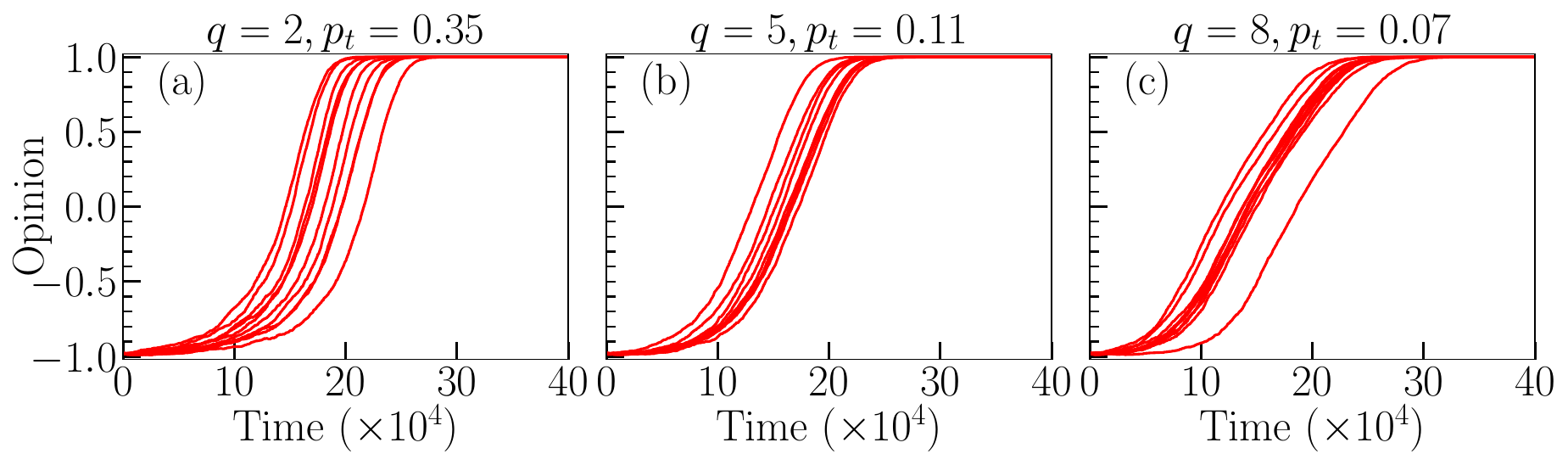}
    \caption{The evolution of the order parameter $m$ at the threshold probability $p_t$ for $q = 2, 5$, and $8$ demonstrates that all samples converge to complete consensus, signifying that, ultimately, every agent aligns with the same opinion. This observation is made with an initial opinion fraction $c_0 = 0.01$ and a population size of $N = 5000$.}
    \label{fig:mcsp1}
\end{figure}

We analyze the evolution of the order parameter $m$ over time, starting from a low initial opinion fraction $c_0 = 0.01$, at the probability threshold $p_t$, and for a population size of $N = 5000$. As shown in Fig.~\ref{fig:mcsp1}, for different scenarios such as $q = 2$ at $p_t = 0.35$, $q = 5$ at $p_t = 0.11$, and $q = 8$ at $p_t = 0.07$, all samples eventually achieve a homogeneous state (complete consensus) with $m = +1$. This outcome indicates that at these threshold probabilities, the model does not exhibit a separation point that would otherwise lead public opinion to diverge into two consensus states with $m= -1$ and $m = +1$.
 
\begin{figure}[tb!]
    \centering
    \includegraphics[width =\linewidth]{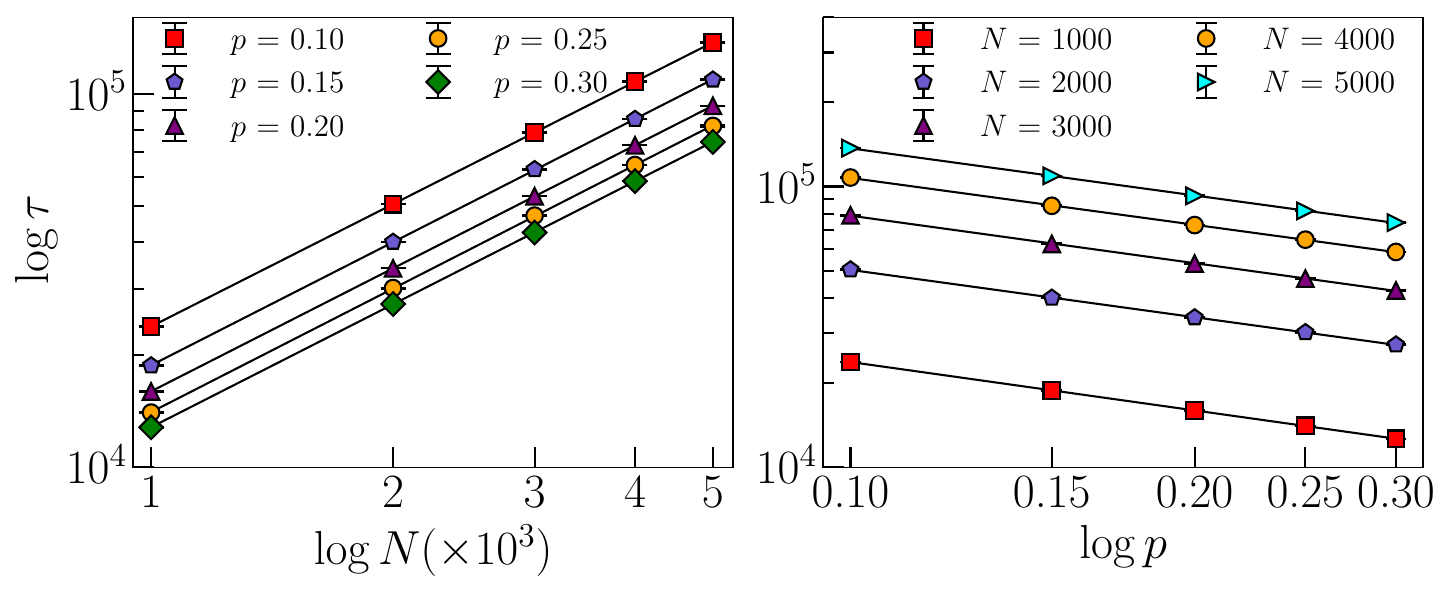}
    \caption{(a) A log-log graph depicting the relationship between the relaxation time $\tau$ and the population size $N$ for the model with $q = 2$ and for typical values of $p \leq p_t$. Its relation is characterized by a power-law $\tau(N) \propto N^{\nu}$, with a slope $\nu \approx 1.093$ representing the trend across all data $p$. (b) A log-log graph showing the relaxation time $\tau$ as a function of the probability $p$, which exhibits a power-law behavior $\tau(p) \propto p^{\delta}$, where $\delta \approx -0.560$ as the slope of all data $N$. The initial opinion is $c = 0.5$, and each point is the average of $10000$ samples.}
    \label{fig:relax}
\end{figure}

Next, we want to analyze the relaxation or consensus time, denoted by $\tau$. This parameter represents the time required for the system to reach a homogeneous state or a complete consensus, wherein all agents adopt the same opinion. We aim to explore the dependence of the relaxation time $\tau$ on the population size $N$. To achieve this point, we vary $N$ values, \textcolor{black}{namely $1000, 2000, 3000, 4000,$ and $5000$,} with each data point being the mean of over $10000$ samples. The initial state is set to be completely disordered ($c_0 = 0.5$), and the probability $p$ not exceeding the threshold $p_t$. The numerical result for $q = 2$ is exhibited in Fig.~\ref{fig:relax} (a), which follows a power-law relation that can be written as:
\begin{equation}\label{eq:relax_log_N}
    \tau (N) \propto N^{\nu},
\end{equation}
where $\nu \approx 1.093$ represents the slope of the data. \textcolor{black}{This value of \( \nu \) is close to that obtained for the voter model on Erdos-Renyi networks \cite{castellano2005comparison, vazquez2008analytical}, the \( q \)-voter model on complete graphs with \( q = 2, 3 \) \cite{moretti2013mean}, and the voter model on \( d \)-lattices with \( d > 2 \) \cite{blythe2010ordering, krapivsky2010kinetic}, all of which have \( \nu = 1 \). Generally, similar power-law relationships have been observed in other opinion dynamics models across various scenarios \cite{castellano2005comparison, sood2005voter, vazquez2008analytical, sousa2008effects, crokidakis2012effects, moretti2013mean, krapivsky2010kinetic, blythe2010ordering, azhari2023external, corberi2024ordering, muslim2024mass, meyer2024time, ramirez2024ordering}.}

For constant \(N\), the log-log plot of the relaxation time \(\tau\) versus the probability \(p\) for \(q = 2\) is depicted in Fig.~\ref{fig:relax} (b). As \(p\) increases, the relaxation time \(\tau\) decreases, indicating an accelerating effect of mass media influence on the system's dynamics. Our analysis shows that the model exhibits a power-law relation between these variables, which can be written as:
\begin{equation}\label{eq:relax_log_p}
    \tau (p) \propto p^{\delta},
\end{equation}
where \(\delta \approx -0.560\). This behavior is similar to our previous study regarding the $q$-voter model on a complete graph \cite{muslim2024mass}. The behavior consistency of the relaxation time across different network topologies suggests a universal aspect of the system's response to mass media influence. The negative value of \(\delta\) indicates that as the probability \(p\) increases, the relaxation time \(\tau\) decreases, thereby speeding up the consensus formation process.

\textcolor{black}{To clarify the results of this study, we summarize the exponent values \( \nu \) and \( \delta \) obtained from various studies on the voter and \( q \)-voter models across different network topologies, as shown in Table~\ref{tab:table1}. These studies show that the value of \( \nu \) is $\nu \in (0,2]$. The smallest value is found in Ref.~\cite{ramirez2024ordering}, with the form \( \tau(N) \propto \ln(N) \), indicating that \( \nu \) is a small positive constant. The largest value, \( \nu = 2 \), is found in Refs.~\cite{krapivsky2010kinetic, blythe2010ordering, meyer2024time}, which have the form \( \tau (N) \propto N^2 \).}

 \begin{table*}
\caption{\label{tab:table1}%
The values of \( \nu \) and \( \delta \) are based on Eqs. \eqref{eq:relax_log_N} and \eqref{eq:relax_log_p} for the voter and \( q \)-voter models on various network topologies under different scenarios. See the Refs. for detailed scenarios.}
\begin{ruledtabular}
\begin{tabular}{lllcr}
 Model & Network Topology & $\nu$ & $\delta$ & Refs. \\
 \hline
Voter & Erdos-Renyi & $1$ & - & \cite{castellano2005comparison,vazquez2008analytical} \\
Voter & Barabasi-Albert & $ \in (0,1)$ & - & \cite{sood2005voter, vazquez2008analytical} \\
\multirow{2}{*}{\textit{Q}-voter}  & \multirow{2}{*}{Complete graph}  & $1$ for $q = 2, 3$ & \multirow{2}{*}{-} & \cite{moretti2013mean} \\
                                   &                                  & $1/2$ for $q = 4$ &  & \cite{moretti2013mean} \\
\textit{Q}-voter & Complete graph & depends on probability $p$ & - & \cite{azhari2023external}\\
\textit{Q}-voter & Complete graph & depends on probability $p$ & depends on $q$ & \cite{muslim2024mass} \\
\multirow{3}{*}{Voter} & \multirow{3}{*}{$D$-lattice}  & $2$ for $D = 1$ & \multirow{3}{*}{-} & \cite{krapivsky2010kinetic, blythe2010ordering}\\
 &  & $ > 1$ for $D = 2$ &  & \cite{krapivsky2010kinetic, blythe2010ordering, corberi2024ordering}\\
  &  & $ = 1$ for $D > 2$ &  & \cite{blythe2010ordering,krapivsky2010kinetic}\\
Noisy voter & Complete graph & $2$ & - & \cite{meyer2024time}\\
\textit{Q}-voter & Complete graph & is a small positive constant & - & \cite{ramirez2024ordering}\\
\textit{Q}-voter & Barabasi-Albert & $1.093$ for $q = 2$ & $-0.560$ for $q = 2$& This study\\
\end{tabular}
\end{ruledtabular}
\end{table*}

Despite the simplicity of this model, understanding these scaling behaviors is crucial for modeling and predicting how quickly a population can reach consensus under varying conditions of mass media influence. This type of modeling can be adapted to reflect existing social conditions, mass media behavior, societal responses to media, and other factors. This knowledge can be applied to optimize strategies in social influence campaigns, ensuring that desired outcomes are achieved efficiently. Whether in the context of political campaigns, public health initiatives, or marketing efforts, leveraging the insights from these models can enhance the effectiveness of efforts to shape public opinion.

\section{Summary}

This study examines the effect of mass media influence on the evolution of opinions within the context of the $q$-voter model on a Barabási-Albert network. The minimum node degree of the Barabási-Albert network is equal to the size of the $q$-sized agent group. Within the population, a group of agents of size $q$ and a voter are randomly selected. The voter will adopt the group's opinion if it is unanimous, or, with probability $p$, the voter will adopt the mass media opinion if it is not unanimous.

We conducted extensive numerical simulations to investigate the impact of mass media on the consensus state of the average public opinion, represented by the order parameter $m$. We identified a threshold probability $p_t$ that invariably causes the system to evolve towards a consensus state. Specifically, for $p \geq p_t$, the system consistently converges to a homogeneous state characterized by $m = 1$, even when the initial proportion of the ``up" opinion is small, namely $c_0 = 0.01$. The threshold probability $p_t$ depends on the group size $q$ and follows a power-law $p_t(q) \propto q^{\alpha}$, where $\alpha \approx -1.187$. This result indicates that the threshold probability decreases as the group size $q$ increases.

We further explored the scaling behavior within this model by employing standard finite-size scaling analysis to ascertain the scaling parameters for the order parameter \( m \) versus the initial opinion \( c_0 \). The separation point delineating the consensus states can be readily determined by identifying the intersection point between the order parameter \( m \) and the initial opinion \( c_0 \). The best scaling parameters obtained for this model are \( \gamma_1 \approx 0.020 \) and \( \gamma_2 \approx 0.498 \), applicable for \( p \leq p_t(q) \). These scaling parameters facilitate collapsing all data points for different \( N \) into a unified framework, in other words, universally applicable for all values of \( N \). These scaling parameter values are also similar to the scaling parameters obtained for the \( q \)-voter model on a complete graph \cite{azhari2023external, muslim2024mass}.

Furthermore, we analyzed the model's relaxation time $\tau$, which denotes the time required for the system to achieve a homogeneous state. Our results for $q = 2$ reveal that the relaxation time $\tau$ increases with the size of the population $N$ and follows a power-law $\tau(N) \propto N^{\nu}$, where $\nu \approx 1.093$. This kind of relationship is similar to the ones found in various studies on the voter and \( q \)-voter models \textcolor{black}{\cite{castellano2005comparison, sood2005voter, vazquez2008analytical, sousa2008effects, crokidakis2012effects, moretti2013mean, krapivsky2010kinetic, blythe2010ordering, azhari2023external, corberi2024ordering, muslim2024mass, meyer2024time, ramirez2024ordering}.} Additionally, the impact of mass media influence reduces the relaxation time, meaning that $\tau$ decreases as the probability $p$ increases. These two parameters also follow a power-law $\tau(p) \propto p^{\delta}$, where $\delta \approx -0.560$, demonstrating that mass media influence significantly accelerates the system's convergence to consensus.

\section*{Author Contributions}
\textbf{R.~F:} Conceptualization, Methodology, Writing, Formal analysis,   Review \& editing. \textbf{Z.~A:} Writing, Formal analysis, Review \& editing. \textbf{R.~M:} Main contributor, Conceptualization, Methodology, Software, Formal analysis, Validation, Writing - original draft, Visualization, Review \& editing, Supervision. 

\section*{Declaration of Interests}
The authors declare that they have no known competing financial interests or personal relationships that could have appeared to influence the work reported in this paper.

\section*{Acknowledgments}
The authors extend their gratitude to Universitas Andalas for its financial support and to the BRIN Research Center for Quantum Physics for granting access to its QuaSi computers, which enabled the performance of the numerical simulations.

\nocite{*}

\bibliography{apssamp}

\end{document}